\documentclass[11pt]{article}
\usepackage[english]{babel}
\usepackage{amsmath,amssymb,graphicx,cite,color,hyperref
}
\usepackage[left=3cm,right=3cm,top=3cm,bottom=3cm,head=1cm,foot=1cm]{geometry}
\usepackage{scrextend}
\usepackage{epstopdf}
\newcommand{\out}{\mathrm{out}}
\newcommand{\side}{\mathrm{side}}

\begin{document}
\title{Higher order anisotropies in the Buda-Lund model: Disentangling flow and density field anisotropies}

\author{S\'andor L\"ok\"os$^1$, M\'at\'e Csan\'ad$^{1,2}$, Boris Tom\'a\v{s}ik$^{3,4}$, Tam\'as Cs\"org\H{o}$^{5,6}$\\
{\small $^1$E\"{o}tv\"{o}s Lor\'{a}nd University, H-1117 Budapest, P\'azm\'any P. s. 1/a, Hungary} \\
{\small $^2$Stony Brook University, Stony Brook, NY, 11794-3400, USA} \\
{\small $^3$Univerzita Mateja Bela, 97401 Bansk\'a Bystrica, Slovakia} \\
{\small $^4$Czech Technical University in Prague, FNSPE, 11519 Prague, Czech Republic} \\
{\small $^5$Wigner RCP, H-1525 Budapest 114, P.O.Box 49, Hungary} \\
{\small $^6$EKU KRC, H-3200 Gy\"ongy\"os, M\'atrai \'ut 36, Hungary}
}

\maketitle

\begin{abstract}
The Buda-Lund hydro model describes an expanding ellipsoidal fireball, and fits the observed elliptic flow and oscillating HBT radii
successfully. Due to fluctuations in energy depositions,
the fireball shape however fluctuates on an event-by-event basis.
The transverse plane asymmetry can be translated into a series of multipole anisotropy coefficients.
These anisotropies then result in measurable momentum-space anisotropies, to be measured with
respect to their respective symmetry planes. In this paper we detail an extension of the Buda-Lund model to multipole
anisotropies and investigate the resulting flow coefficients and oscillations of HBT radii.
\end{abstract}

\section{Introduction}
\label{s:intro}

Anisotropies in distributions of hadrons produced in ultrarelativistic heavy-ion 
collisions are the key observable in the quest for the transport properties of the 
hot strongly interacting matter. Through a careful comparison of measured data with 
theoretical predictions access is open to the values of shear and bulk viscosity, 
the equation of state, equilibration time, and other properties of the matter. 
Distributions of hadrons are formed at the very last moment of fireball history
when all hadrons leave from the hot and strongly coupled system. It is therefore 
instructive to understand how the observed anisotropies of the momentum 
distribution are connected to the shape and the expansion pattern of the fireball 
at freeze-out. Helpful tools for this task are the hydrodynamically inspired 
models of hadron production which allow for easy simulation of different 
final states of the fireball. Here we will use a model that is in close connection to exact
solutions of hydrodynamics as well---the Buda-Lund model~\cite{Csorgo:1995bi,Csanad:2003qa}.

There are two kinds of source anisotropies which are translated into the anisotropy 
of hadron distributions. One is connected with the shape of the fireball and the 
other with the angular dependence of its expansion velocity. It has been 
investigated in the past how they both individually contribute to the second-order 
anisotropy of single-particle distributions and oscillations of correlation radii
in femtoscopy~\cite{Csanad:2008af,Ster:2010ia}. 

Unprecedented statistics collected in nuclear collisions at the LHC and at RHIC
allow also detailed study of higher-order
anisotropies and higher-order oscillations of Bose-Einstein correlation radii~\cite{Adare:2011tg,Aamodt:2011by,Adamczyk:2013waa,Adare:2014vax}.
There is at present no sufficient 
theoretical understanding of the femtoscopic measurements. 

To this end, we extend here the Buda-Lund model so that it includes the third-order 
anisotropy in shape and expansion velocity. Then we give an outlook at the 
extension to even higher orders. We further calculate the third-order 
angular dependence of the spectra and correlation radii and analyze how 
it is influenced by the different features of the model. 

The model provides description of direct hadron production without the inclusion of
resonance decays. Although resonances are not included in our current investigation, we note that
the core-halo model was developed to correct the hydrodynamical and
phenomenological calculations, similar to the ones presented here, to take into
account the effects of decays from long-lived resonances~\cite{Csorgo:1994in}.
Long lived resonances decay in a halo region, characterized by large
length-scales (typically 20 fm or larger) that are outside the hydrodynamically
evolving, hot and dense hadronic matter, that is referred to as the  core
region.  Comparing the core-halo model calculations to experimental data, the
main effects of resonances appear to be the modification of the single-particle
spectra of pions, and the suppression of the strength of the two-pion
Bose-Einstein correlations, however, their effects on the short range
Bose-Einstein correlation or HBT radii and on the anisotropies of the single
particle spectra are negligible, as indicated by the hydrodynamical scaling of
these observables~\cite{Csorgo:1999sj,Csanad:2003qa,Csanad:2005gv}.


\section{The Buda-Lund model}
\label{s:model}

The Buda-Lund model is formulated in terms of the source function, which represents the 
(Wigner) probability density
of particle creation at a space-time point $x=(t,r_x,r_y,r_z)$ and  four-momentum $p=(E,p_x,p_y,p_z)$.
It generally takes the form of a J\"uttner-type statistical distribution:
\begin{align}
S(x,k)d^4x=\frac{g}{(2\pi)^3}\frac{p^\mu d^4\Sigma_\mu(x)}{B(x,p)+s_q}
\label{eq:source_function}
\end{align}
where $g$ stands for  degeneracy factor, $p^\mu d^4\Sigma_\mu$ is the Cooper-Frye factor~\cite{Cooper:1974mv},
$s_q$ is  a
quantum-statistical term, being $-1$ for Bose-Einstein, 1
for Fermi-Dirac statistics, and 0 for Maxwell-Boltzmann statistics,
and the thermodynamic distribution  $B(x,p)$ takes the form
\begin{align}
B(x,p)=\exp\left [ \frac{p_\mu u^\mu(x)}{T(x)}-\frac{\mu(x)}{T(x)} \right ]\, .
\end{align}
The freeze-out in this model happens along the hypersurface perpendicular to 
the flow velocity and the  Cooper-Frye factor is expressed as \cite{Csanad:2003qa}:
\begin{align}
p^\mu d^4\Sigma_\mu(x) = p^\mu u_\mu(x) H(\tau)d^4x
\end{align}
where $H(\tau)$ is the freeze-out probability density in proper time, with $\tau$ being the proper 
time in the local frame co-moving with the velocity $u_\mu$. 
The smearing factor of the freeze-out time  will be assumed to take the form of a delta function:
$H(\tau)=\delta(\tau-\tau_0)$. This simplification corresponds to the approximation when the
freeze-out happens suddenly at $\tau_0$ freeze-out time.

The velocity field is  calculated from a potential $\Phi(x)$:
\begin{equation}
u_\mu = \gamma(1,{\bf v})=\gamma(1, \partial_x \Phi, \partial_y \Phi, \partial_z \Phi)
\label{eq:velocity}
\end{equation}
where $\gamma = 1/\sqrt{1-(\partial_x\Phi)^2-(\partial_y\Phi)^2-(\partial_z\Phi)^2}$. 
With the potential field we shall be able to introduce azimuthal angle variations of the expansion 
velocity field.
 
The model includes gradients in temperature profile:
\begin{align}
\frac{1}{T(x)}=\frac{1}{T_0}\left(1+a^2 s\right)
\end{align}
where $a$ parametrizes the gradient and $s$ is a scaling variable which depends on spatial coordinates
and will be specified later. 
The parameter $a^2$ can also  be expressed as
\begin{align}
a^2=\frac{T_0-T_s}{T_s}
\end{align}
where $T_0$ is the central temperature and $T_s$ is the temperature at the surface of the fireball.
The scaling variable $s$ is important when the parametrization is identified as a solution of a certain class 
of hydrodynamic models~\cite{Csorgo:2003ry,Csanad:2014dpa}. In such a case, its co-moving 
derivative must vanish
\begin{equation}
\label{e:ders}
u^\mu \partial_\mu s = 0\,  .
\end{equation}

The fugacity term is defined similarly
\begin{align}
\frac{\mu(x)}{T(x)}=\frac{\mu_0}{T_0}-bs
\end{align}
i.e.\ the parameter $b$ is the density gradient. (Note that $b=1$ was assumed in earlier versions of the
Buda-Lund model.)

Next we extend the previous formulation of the Buda-Lund model so that anisotropy 
in azimuthal angle up to third order is included. 

There are two kinds of asymmetries which we investigate in this paper: the spatial asymmetry and 
the velocity field asymmetry.
The former can be described by the scale variable $s$. For completeness and for the sake 
of example let us review that the perfectly symmetric case (not investigated here) would be 
that of spherical symmetry with 
\begin{equation}
s=\frac{r_x^2+r^2_y + r^2_z}{R_\circ^2}\label{e:sphericals}
\end{equation}
and the spheroidal symmetry with distinguished longitudinal direction
\begin{equation}
s = \frac{r^2}{R^2} + \frac{r_z^2}{Z^2}\,  ,\label{e:spheroids}
\end{equation}
with $r^2 = r_x^2 + r_y^2$. Depending on the model, $R_\circ$ and/or $R$ is the radial scale 
and $Z$ is the longitudinal 
scale. Ellipsoidal symmetry would be then represented by
\begin{equation}
s=\frac{r_x^2}{X^2}+ \frac{r^2_y}{Y^2} +  \frac{r_z^2}{Z^2}.\label{e:ellipsoidals}
\end{equation}

The anisotropy in transverse shape can be introduced for the elliptic deformation (second order) also like 
\begin{equation}
s=\frac{r^2}{R^2}\left(1+\epsilon_2\cos(2\alpha) \right)+\frac{r_z^2}{Z^2} \, ,\label{e:eps2s}
\end{equation}
with $\cos\alpha=r_x/r$, and a connection to the transverse principal axes of the ellipsoid ($X,Y$)
can be established via $X=R/\sqrt{1+\epsilon_2}$ and $Y=R/\sqrt{1-\epsilon_2}$. A triangular deformation
(third order) can also be introduced as
\begin{equation}
s=\frac{r^2}{R^2}\left(1+\epsilon_3\cos(3\alpha) \right)+\frac{r_z^2}{Z^2}\,  .
\end{equation}
The extension to asymmetry of arbitrary order is straightforward
\begin{equation}
s = \frac{r^2}{R^2}\left(1+\sum_n \epsilon_n \cos\left (n(\alpha-\alpha_n)\right )\right)
+\frac{r_z^2}{Z^2}\,  ,\label{e:generals}
\end{equation}
similarly to Ref.~\cite{Csanad:2014dpa}. The phase factors  
$\alpha_n$ must generally be included to reflect the different orientation of
the $n$th order event planes. Their influence will be investigated in more detail
in Section~\ref{s:rad}.

The expansion velocity field is obtained via eq.~(\ref{eq:velocity}) from the potential $\Phi(x)$.
We generalize now the parametrization for $\Phi(x)$ in a similar way as we did for the scaling 
variable.  Let us connect with previous works by recalling that for the spherically symmetric case one 
chooses
\begin{align}
\Phi(x) =\frac{H_\circ}{2}r^2,\;
	{\bf v}=\left (H_\circ r_x,H_\circ r_y, H_\circ r_z \right )
\end{align}
where $H_\circ$ is the radial Hubble-parameter, and eq.~(\ref{e:ders}) with $s$ from (\ref{e:sphericals})
can be fulfilled via the choice of
$H_\circ=\frac{\dot R_\circ}{2R_\circ}$. The velocity profile with 
ellipsoidal expansion symmetry has so far been param\-e\-trized via \cite{Csanad:2003qa} 
\begin{align}
\Phi(x)=\frac{H_x}{2}r_x^2+\frac{H_y}{2}r_y^2+\frac{H_z}{2}r_z^2,\\
	{\bf v}=\left (H_x r_x,H_y r_y, H_z r_z \right )\, ,
\end{align}
where $H_{x,y,z}$ are the directional Hubble-parameters, and eq.~(\ref{e:ders}) with $s$ from (\ref{e:ellipsoidals})
can be fulfilled via the choice of $H_x=\frac{\dot X}{2X}$ and similarly for $y$ and $z$. Here
$X$, $Y$, $Z$ are the length scales in three perpendicular directions and 
$\dot X$, $\dot Y$, $\dot Z$ their proper time derivatives. 
Here we shall use  a form that is 
more straightforwardly generalized to anisotropies of higher orders. The elliptic anisotropy 
is  then param\-e\-trized as
\begin{align}
\Phi(x)=&\frac{H}{2}r^2\left(1+\chi_2\cos(2\alpha) \right) + \frac{H_z}{2}r_z^2 
\end{align}
where $H$ is the radial Hubble-parameter, while $H_z$ is the one describing longitudinal expansion.
This form fulfills eq.~(\ref{e:ders}) with $s$ from eq.~(\ref{e:eps2s}), if
\begin{align}
\dot \epsilon_2  = -2H\chi_2(1-\epsilon_2^2)\, ,\; \frac{\dot R}{R}=H(1-\epsilon_2\chi_2)\, ,\;\frac{\dot Z}{Z}=H_z.
\end{align}
For arbitrary-order asymmetries, $\Phi$ can be introduced via a general form 
\begin{equation}
\label{eq:genphi}
\Phi(x)=\frac{H}{2}r^2\left( 1+\sum_{n=2}^\infty \chi_n \cos (n(\alpha-\alpha_n))\right)+ \frac{\dot Z}{2Z}r_z^2\, .
\end{equation}
With this choice of $u^\mu$ and $s$ from eq.~(\ref{e:generals}),  eq.~(\ref{e:ders}) can be satisfied if one requires
$\chi_n=0$ and the time derivative of $\epsilon_n$ to vanish. This yields a Hubble-like
expansion without the change in the asymmetries. 

\begin{figure}
\centering
\includegraphics[width=0.67\linewidth]{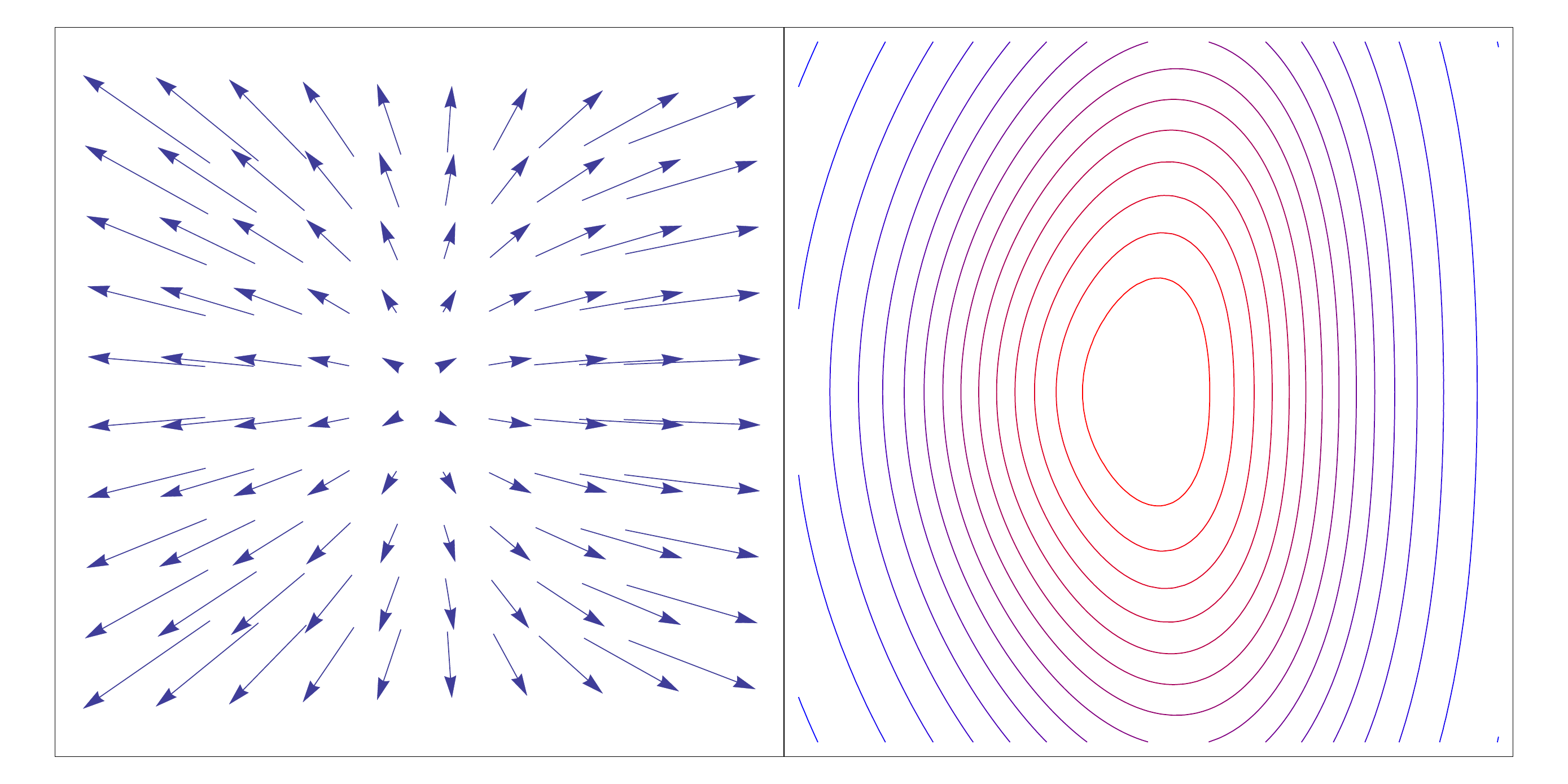}
\caption{\label{fig:fields}Example flow (left) and density (right) fields with example values of
$\epsilon_2=\chi_2=0.3$ and $\epsilon_3=\chi_3=0.1$ for the asymmetry coefficients.}
\end{figure}

More generally, eq.~(\ref{e:ders}) can also be fulfilled, if $\epsilon_n$ and $\chi_n$ are so small that 
their bilinear and quadratic terms can be neglected. In this case,
\begin{align}
\frac{\dot R}{R}=H\;\textrm{ and }\;
\dot\epsilon_n=-2\frac{\dot R}{R}\chi_n
\end{align}
are required to fulfill eq.~(\ref{e:ders}).
The complete set of conditions that are derived from eq.~(\ref{e:ders}) for the $\alpha_n=0$ case includes:
\begin{align}
\frac{\dot{R}}{R}&=H\left(1+\frac{1}{2}\sum_{n=1}^\infty\epsilon_n \chi_n\left(1+\frac{n^2}{4}\right) \right),
\end{align}
as well as for any $k>0$
\begin{align}
\dot{\epsilon}_k = 2H \chi_k -2\left(\frac{\dot{R}}{R}-H\right) \epsilon_k
&+ H\sum_{n=1}^\infty \epsilon_n\chi_{n+k} \left(1+\frac{n(n+k)}{4}\right) \\\nonumber
&+H \sum_{\substack{n=1 \\ n\neq k}}^\infty \epsilon_n \chi_{|n-k|} \left(1+\frac{n(n-k)}{4}\right).
\end{align}
If only a subset of $\epsilon_n$ and $\chi_n$ values are nonzero, then this set of equations can be solved successively.
Let us recall that since the co-moving derivative of $s$ vanishes in all the above cases, 
the Buda-Lund profile described above can be taken as a result of a hydrodynamic calculation, and may form
the basis of exact hydrodynamic solutions. 
Note, however, that this will \emph{not} be required in the present study.
For an illustration of the density and flow fields defined above, see Fig.~\ref{fig:fields}.


\section{Single particle distributions}
\label{s:1part}

Observable quantities, like invariant momentum distributions or the correlation radii, 
are generally obtained by integrating and calculating the space-time moments of the
source function, eq.~\eqref{eq:source_function}. If this cannot be accomplished analytically 
then numeric methods have to be resorted to.


The invariant momentum distribution is obtained as 
\begin{align}
E\frac{d^3N}{dp^3} = N_1(p) = \int d^4x\, S(x,p)\, .
\end{align}
In this paper we shall concentrate on midrapidity, as is appropriate for the description 
of the experiments at RHIC and the LHC. Furthermore, when looking at azimuthally integrated
distributions we shall have to integrate over the azimuthal angle of the emitted hadrons
(which we denote by $\phi$ here)
\begin{equation}
\frac{d^2N}{p_t\, dp_t\, dy} = \int_0^{2\pi} d\phi\, N_1(p)\,  .
\end{equation}

The anisotropies of the transverse momentum distribution are denoted by $v_n$.
These observables drew much experimental as well as theoretical interest recently,
and are defined as $n$-th order Fourier coefficients with respect to the 
angle of the $n$-th order event plane $\psi_n$:
\begin{align}
\frac{\frac{d^3N}{p_t\, dp_t\, dy \, d\phi}}{\frac{d^2N}{2\pi p_t\, dp_t\, dy} } = 
1+2\sum_{n=1}^\infty v_n\cos (n(\phi-\psi_n)).
\end{align}
The anisotropy coefficients $v_n$ can be obtained from 
\begin{equation}
v_n e^{i n \psi_n} 
= 
\frac{\int_0^{2\pi} d\phi \, e^{in\phi}\, N_1(p)}
{\int_0^{2\pi} d\phi \,  N_1(p)}\,  .
\label{e:vndef}
\end{equation}
Usually, $v_2$ is referred to as elliptic flow and $v_3$ as triangular flow.



In our calculations we chose the values of parameters that are motivated by earlier 
fits to data~\cite{Csanad:2004mm,Ster:2010ia}. They are 
listed in Table~\ref{tab:params}.
\begin{table}
\centering
\begin{tabular}{lcc}
\hline
\hline
Particle mass  				&	$m$			&	140 MeV	\\
Freeze-out time 					&	$\tau_0$		&	7 fm$/c$\\
Central freeze-out temperature		&	$T_0$		&	170 MeV\\
Temperature-asymmetry parameter	&	$a^2$		&	0.3 \\
Spatial slope parameter			&	$b$			&	--0.1	\\
Transverse size of the source		&	$R$			&	10 fm\\
Longitudinal size of the source		&	$Z$			&	15 fm\\
Transverse expansion				&	$\dot R$		&	1\\
Longitudinal expansion			&	$\dot Z$		&	0.94\\
\hline\hline
\end{tabular}
\caption{Default values of the model parameters (at the freeze-out). Anisotropy 
parameters are not listed here because they are varied in the different studies 
reported here. }\label{tab:params}
\end{table}

\begin{figure}
\centering
\includegraphics[angle=-90,width=0.5\linewidth]{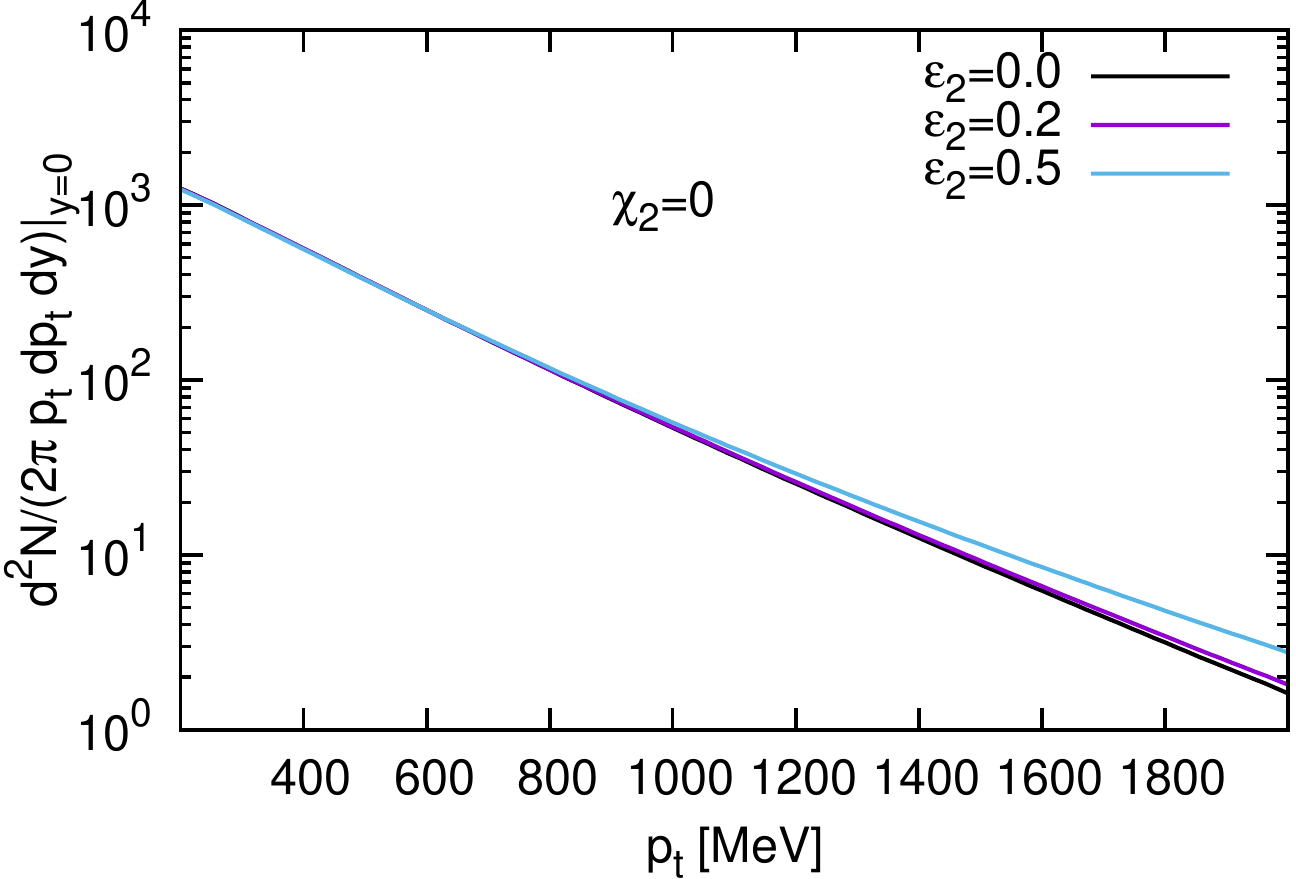}
\caption{\label{fig:N1} Azimuthally integrated single-particle $p_t$ spectra for 
various values of the second-order anisotropy parameter $\epsilon_2$.}
\end{figure}

Generally, one expects that the azimuthal anisotropy has no influence on the azimuthally
integrated spectrum. However, the space and flow anisotropies in the present model 
modify the effective volume, particularly at high $p_t$ where the anisotropy is more 
pronounced. This is illustrated in Fig.~\ref{fig:N1}.
Thus there is a slight flattening of the single-particle $p_t$ spectra connected with the 
increase of the anisotropy parameters $\epsilon_i$ and/or $\chi_i$. Note that in the 
figure we only show the dependence on $\epsilon_2$ but this is qualitatively similar 
to the dependences on other parameters. 

\begin{figure}
\centering
\includegraphics[angle=-90,width=0.67\linewidth]{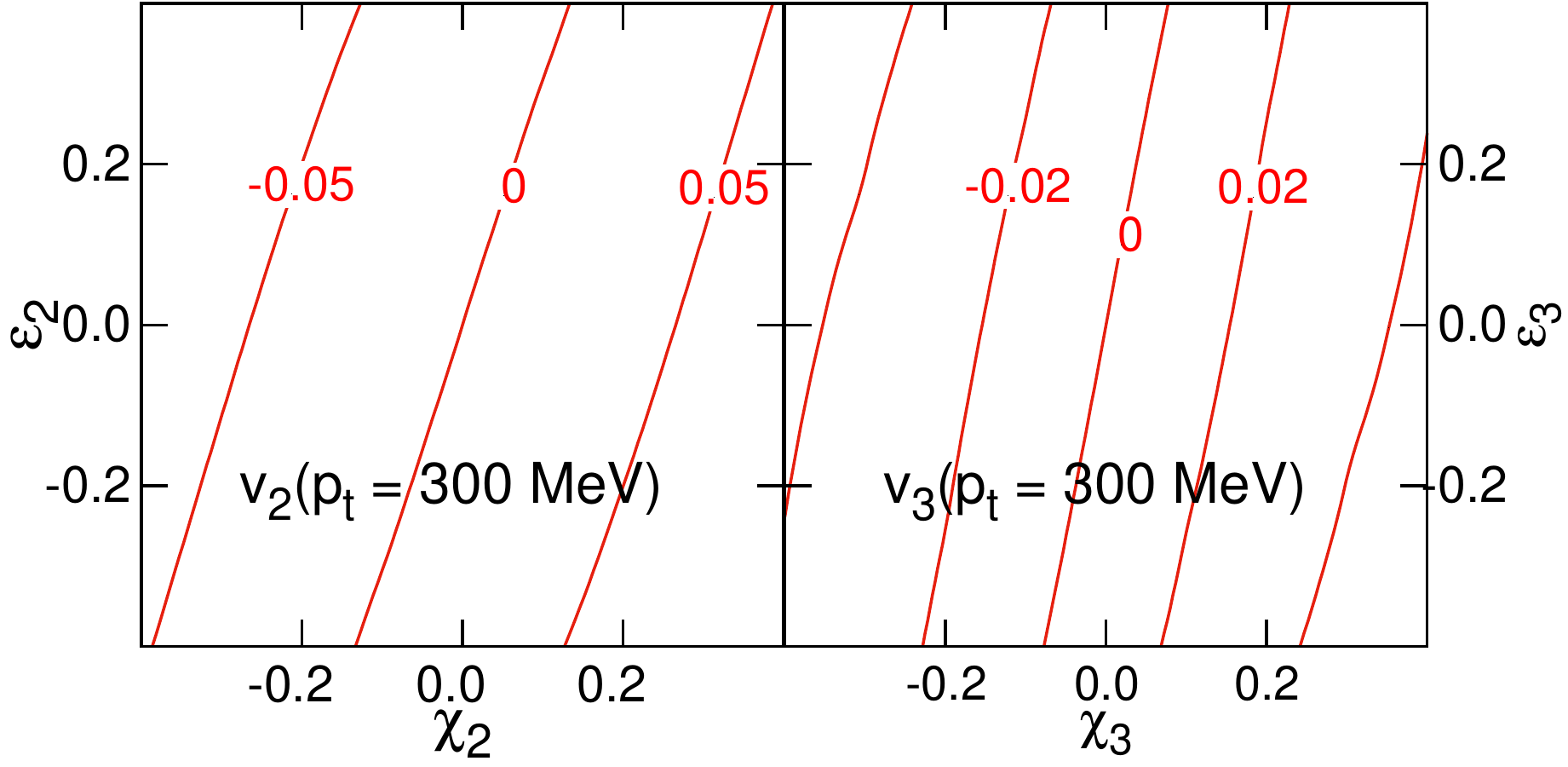}
\caption{\label{fig:flows} The effect of various anisotropy parameters on the flow coefficients.
at $p_t = 300$~MeV/$c$.}
\end{figure}

It has been calculated in~\cite{Csanad:2008af} how the elliptic flow coefficient 
$v_2$ depends on the second order anisotropy in shape and expansion. Here, in
Fig.~\ref{fig:flows} 
we again plot this dependence together with the dependence of $v_3$ on 
$\epsilon_3$ and $\chi_3$. Similarly to the second order, also here we have an
ambiguity: same values of $v_i$ can be obtained from various combinations of 
$\epsilon_i$ and $\chi_i$. Note that $\epsilon_2$ and $\chi_2$ do not
influence $v_3$, and vice versa.


\section{Correlation radii}
\label{s:rad}

The correlation radii are very important quantities for the  exploration of 
the space-time structure of the source. Generally, the two particle momentum correlation function is defined as
\begin{align}
C_2(p_1,p_2)=\frac{N_2(p_1,p_2)}{N_1(p_1)N_1(p_2)}.
\end{align}
The denominator which provides the two-particle distribution with no correlations is 
experimentally obtained by means of event mixing. Usually one introduces 
the momentum difference $q$  and the average pair momentum $K$
\begin{align}
q =p_1-p_2\;\textnormal{ and }\;K = \frac{p_1+p_2}{2}\,  .
\end{align}
Then the correlation function can be determined (within 
reasonable approximation) from the source function 
\begin{align}
C_2(q,K)=1+ \frac{\left | \int d^4x \,  e^{iqx} {S}(x,K)\right |^2}%
{\left ( \int d^4x {S}(x,K) \right )^2} .
\label{eq:corr_func}
\end{align}

If the shape of the correlation function is reasonably close to a Gaussian, then the correlation function 
can be parametrized by a Gaussian
\begin{align}
C(q,K) = 1 + \lambda \exp \left ( - \sum_{i,j} R^2_{ij}\, q_i\, q_j \right ) 
\end{align}
run over the spatial directions, and where the parameter $\lambda$ stands for the
(in general, mean momentum dependent) intercept parameter, that measures the
strength of two-particle Bose-Einstein correlation functions. When particle identification
errors are negligibly small, this parameter $\lambda$ can be interpreted in
the core-halo picture as the (momentum dependent) squared fraction of particles
coming from the hydrodynamically evolving core~\cite{Csorgo:1994in}.

The temporal component of 
$q$ is suppressed via the on-shell constraint
\begin{align}
\label{e:onshell}
q\cdot K = 0 \Rightarrow q^0 = \frac{\vec q\cdot \vec K}{K^0} = \vec q \cdot \vec \beta\, .
\end{align}
For nearly Gaussian, for example hydrodynamically evolving sources, the correlation radii $R_{ij}^2$
can be expressed through spatio-temporal (co)variances of the hydrodynamically evolving, core
part of the source function. The temporal admixture in these radii is due to the
on-shell constraint (\ref{e:onshell}).
Note, however, that long lived resonances typically dominate the variances of
the source distribution even if their decay products are present in a small
relative fraction.  Due to this reason, the evaluation of the correlation radii has to be restricted 
to the core or the hydrodynamically evolving part of the source~\cite{Csorgo:1999sj}.

When studying the azimuthal dependence of the correlation radii, meticulous bookkeeping of the angular 
variables is requested. Recall that we denote that
\begin{labeling}{mi}
\item[$\phi$] is the azimuthal angle of the emitted particles
\item[$\psi_n$] is the $n$-th order event plane determined for the distribution of produced hadrons 
according to eq.~(\ref{e:vndef})
\item[$\alpha$] is the spatial coordinate azimuthal angle  
\item[$\alpha_n$] is the phase of the spatial azimuthal dependence of the source function.
\end{labeling}
It is also useful to notice that $\phi$ and $\psi_n$ are measurable while $\alpha$ and $\alpha_n$ only 
appear in the calculations and cannot be directly accessed by measurement. 

In general, the correlation radii
measure the lengths of homogeneity~\cite{Makhlin:1987gm}. These are sizes of the 
homogeneity regions from which particles with given momentum are produced.
For hydrodynamically expanding fireballs, these homogeneity regions are typically
smaller than the whole volume of the fireball, if the fireball has gradients in the
flow velocity distribution, that lead to variations in the local flow velocities 
that are larger than what can be overcompensated by the locally thermalized velocity
distribution of the emitted particles.

The spatio-temporal distribution of the particle emitting source is frequenty analyzed
in the Bertsch-Pratt side-out-longitudinal decomposition, as measured in the Longitudinal
Center of Mass System of the particle pair (LCMS). LCMS is the frame where the longitudinal
component of a given particle pair has vanishing mean momentum along the beam direction,
the transverse momentum components being the same as in the laboratory. In this frame,
the direction of the mean momentum of a given particle pair defines the outwards or
{\it out} direction, which is perpendicular to the beam direction,  which in turn is
referred to as the longitudinal or {\it long} direction.  The sidewards or
{\it side} direction is perpendicular to both the {\it out } and the {\it long} direction,
so that the ({\it side},{\it out},{\it long}) directions form a right-handed coordinate system.

When analyzing the correlations of particles 
emitted under different azimuthal angles, one looks at the fireball from those angles. This change of 
the viewpoint introduces the \emph{explicit} azimuthal angle dependence of the correlation radii. 

The homogeneity regions change for particles emitted under different azimuthal angles. This introduces the 
\emph{implicit} azimuthal angle dependence of the correlation radii. 

It is instructional to write out the outward and sideward coordinates as 
\begin{subequations}
\begin{align}
r_\out &= r \cos(\alpha - \phi)\\
r_\side & = r \sin(\alpha - \phi)
\end{align}
\end{subequations}
where $\phi$ is defined by the direction of the particle. With this notation we obtain
\begin{subequations}
\begin{align}
R_\out^2(K) &= \left \langle ( r_\out - \beta_t t)^2 \right \rangle_c  - 
\left \langle r_\out - \beta_t t \right \rangle^2\\\label{eq:rside}
R^2_\side(K) &= \left \langle r_\side^2 \right \rangle_c  - \left \langle r_\side \right \rangle_c ^2\,  .
\end{align}
\end{subequations}
where $\beta_t$ is the transverse component of $\beta$ introduced in eq.~(\ref{e:onshell}),
and we have introduced averaging over the source function of the hydrodynamically evolving core
\begin{equation}
\left \langle f(x) \right \rangle_c  = \frac{\int d^4x\, f(x) \, S_c(x,K)}{\int d^4x\, S_c(x,K)}\,  ,
\end{equation}
similarly to e.g. the notation of eqs.~(106)-(110) of ref.~\cite{Csorgo:1999sj}).
Using this method, we have evaluated the correlation radii numerically as
functions of $\phi$ for various azimuthal anisotropy parameters. 

In the real experiment the shape and expansion pattern of the fireball fluctuate from event to event. 
Even if we fix the average transverse size and the anisotropy parameters $\epsilon_n$ and $\chi_n$
there still remain the phases $\alpha_n$ which are unlikely to be correlated for the second and the third order. 
In an experimental analysis one usually rotates all events so that they are aligned according to $\psi_2$ 
or $\psi_3$. By rotating and summing up a large number of events only oscillations of the same order 
(and its multiplicatives) remain as that of the angle of the reaction plane. 

In order to see both the second-order and the third-order oscillations in data one would have to refrain 
from averaging over a large number of events which all have different $\Delta\psi_{23} = \psi_2-\psi_3$. 
Perhaps a way to select events for such an analysis can be provided by the recently proposed Event
Shape Sorting~\cite{Kopecna:2015fwa}. Here we want to investigate what is actually the 
effect of averaging on the $\phi$-dependence of the correlation radii. To this end, we fix the anisotropy parameters
and perform several calculations where we always set the difference of $\alpha_2$ and $\alpha_3$ at
a different value. Note that in the calculation we rotate the source by choosing $\alpha_2$  and $\alpha_3$, 
while the experimental data analysis is done with the event planes $\psi_2$ and $\psi_3$. In a soft particles 
emitting source without resonance decays---like here---those two kinds of directions actually must agree. 

\begin{figure}
\centering
\includegraphics[angle=-90,width=0.6\linewidth]{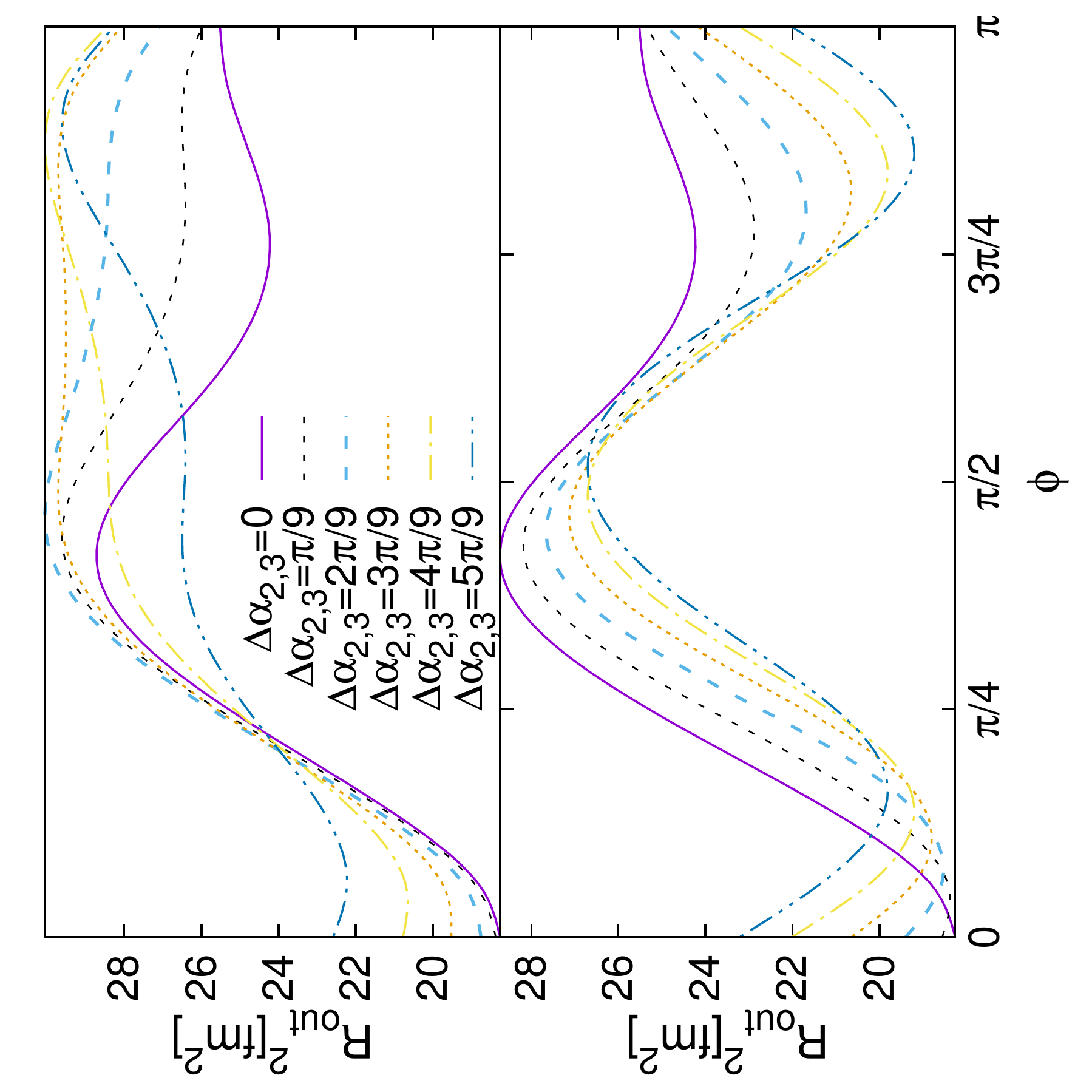}
\caption{\label{fig:rs_angles} 
The $\phi$-dependence of $R_\out^2$ for sources with $\chi_2=0.2$, $\chi_3=0.3$, and 
$\epsilon_2=\epsilon_3=0$. Sources  are oriented so that $\alpha_2 = 0$ (top) and 
$\alpha_3=0$ (bottom). Different curves correspond to different values of 
$\Delta\alpha_{23} = \alpha_2 - \alpha_3$. 
}
\end{figure}

The results for $R_\side^2$ vs. $\phi$ with different 
$\Delta\alpha_{23} = \alpha_2 - \alpha_3$ values are plotted in Fig.~\ref{fig:rs_angles}. 
We clearly see that the gross shape of the dependence is set by the choice of the alignment 
angle. This behavior is best understood qualitatively if we write out the velocity field for 
both cases. If we rotate the source to the second-order event-plane,
i.e.~ $\alpha_2 = 0$, then the transverse velocity field derived from  Eq.~(\ref{eq:velocity}) with 
$\Phi(x)$ given by Eq.~(\ref{eq:genphi}) becomes
\begin{subequations}
\begin{align}
v_x &=  r\frac{\dot R}{R}  \Bigl ( \cos\alpha+\chi_2\cos\alpha
+\frac{\chi_3}{4}(5\cos(2\alpha-\Delta \alpha_{23}) -\cos(4\alpha-\Delta \alpha_{23}))\Bigr ) \\
v_y &=  r\frac{\dot R}{R} \Bigl ( \sin\alpha-\chi_2\sin\alpha
-\frac{\chi_3}{4}(5\sin(2\alpha-\Delta \alpha_{23})+\sin(4\alpha-\Delta \alpha_{23}))\Bigr )\,  .
\end{align}
\end{subequations}
The scaling variable for non-vanishing $\epsilon_2$ and $\epsilon_3$ would be
\begin{equation}
s = \frac{r^2}{R^2}(1+\epsilon_2\cos(2\alpha)+\epsilon_3\cos(3\alpha-\Delta \alpha_{23}))+\frac{r_z^2}{Z^2}\, .
\end{equation}
If, on the other hand, we choose $\alpha_3=0$, then the velocity field becomes
\begin{subequations}
\begin{align}
v_x &=  r\frac{\dot R}{R} \Bigl ( \cos(\alpha)+\chi_2\cos(\alpha+\Delta\alpha_{23})
+\frac{\chi_3}{4}(5\cos(2\alpha)-\cos(4\alpha))\Bigr )  \\
v_y &=  r\frac{\dot R}{R}
\Bigl ( \sin(\alpha)-\chi_2\sin(\alpha+\Delta \alpha_{23})
-\frac{\chi_3}{4}(5\sin(2\alpha)+\sin(4\alpha))\Bigr ) \,  ,
\end{align}
\end{subequations}
and the scaling variable
\begin{equation}
s = \frac{r^2}{R^2}(1+\epsilon_2\cos(2\alpha+\Delta\alpha_{23})+\epsilon_3\cos(3\alpha))+\frac{r_z^2}{Z^2}.
\end{equation}
Due to the mechanism of how the flow velocity is set by Eq.~(\ref{eq:velocity}) the angle difference 
$\Delta\alpha_{23}$ is combined in the two cases with different orders of harmonic oscillations and there 
is no simple shift from one alignment to the other. 

\begin{figure}
\centering
\includegraphics[angle=-90,width=0.6\linewidth]{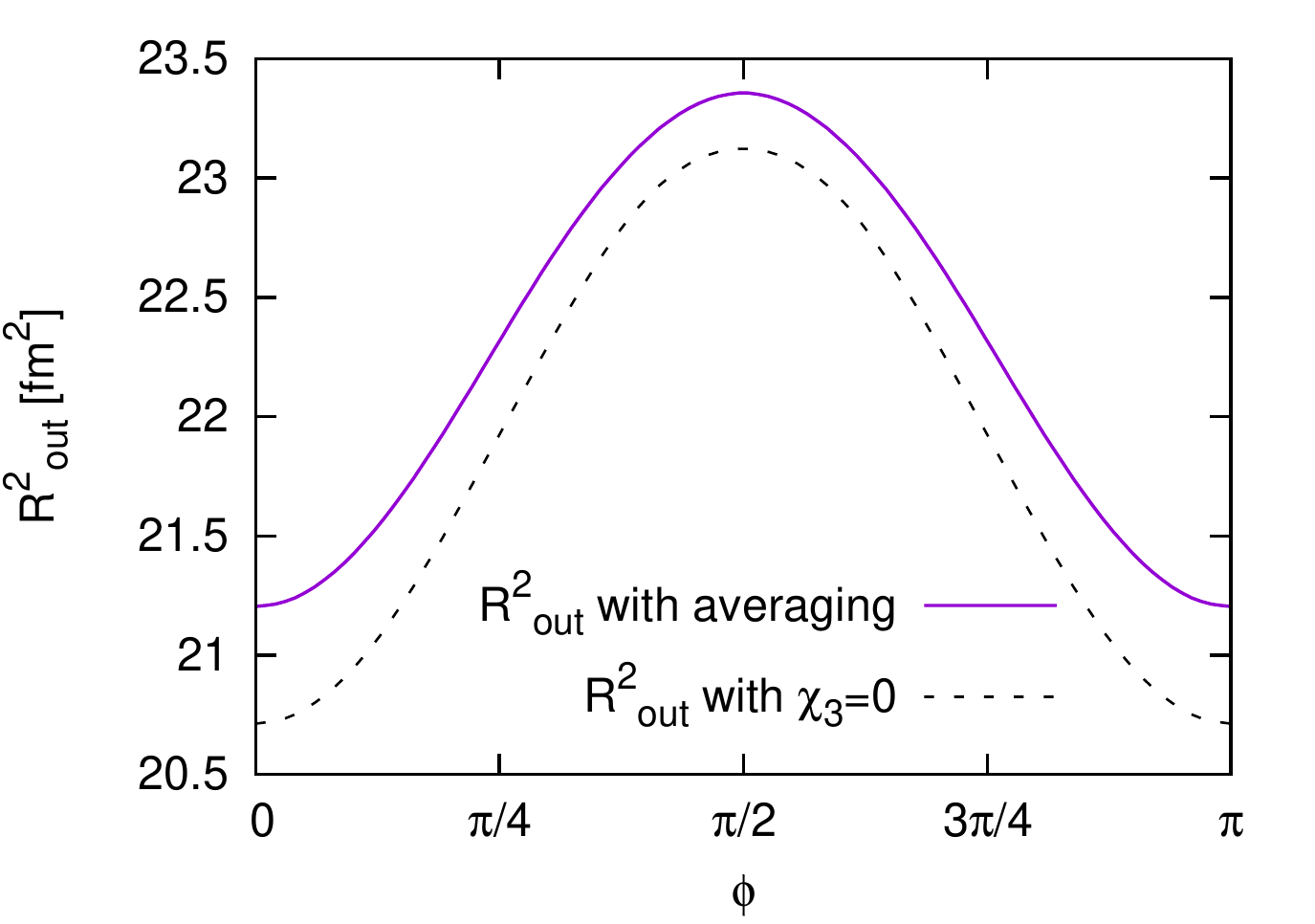}
\caption{\label{fig:raver} 
The $\phi$-dependence of $R_\side^2$ for $\chi_2 = 0.2$, $\epsilon_2 = \epsilon_3 = 0$. 
Solid: $\chi_3 = 0.3$ and the curve is averaged over $\Delta\alpha_{23}$; 
dashed: $\chi_3 = 0$ and no averaging. 
}
\end{figure}

When summing up over a large number of events the various curves are all being averaged into one. 
We want to see the influence of such averaging. To this end, we set $\alpha_2 = 0$ and calculate 
$R_\side^2(\phi)$ in two different ways. First, calculation via Eq.~(\ref{eq:rside}) is performed with 
$\chi_2 = 0.2$ and $\chi_3 = 0.3$ at multiple values of $\Delta\alpha_{23}$ and then the results 
are averaged over $\Delta\alpha_{23}$. Second, $\chi_3$ is set to 0 and the calculation with only the 
second-order parameter $\chi_2 = 0.2$ is performed. Both results are plotted in Fig.~\ref{fig:raver}. 
We observe that the averaging basically preserves the shape of the dependence but it increases 
its absolute size by a relatively small amount. Qualitatively, similar results are observed for averaging over
any of $\epsilon_2$, $\epsilon_3$, $\chi_2$, $\chi_3$. 

This is best understood qualitatively by means of a very simplified model in which, however, 
we keep both the second and the third order variation. Let us write the emission function as
\begin{align}
S_{\mathrm{toy}}(x) &= {\rm e}^{-s} = \exp\left [-\frac{r^2}{R^2}(1 + \epsilon_2 \cos 2\alpha + \epsilon_3 \cos 3(\alpha - \Delta\alpha_{23}))
\right ]\,  .
\end{align}
Then if we average over $\Delta\alpha_{23}$, we get
\begin{align}
S_{\mathrm{toy,av}}(x)&= \langle S_{\mathrm{toy}}(x) \rangle_{\Delta\alpha_{23}} 
= \exp\left[-\frac{r^2}{R^2}(1 + \epsilon_2 \cos 2\alpha)\right] I_0\left(\epsilon_3 \frac{r^2}{R^2}\right)
\end{align}
with $I_0$ denoting the zeroth order modified Bessel function. If we then integrate over $\alpha$, we get
\begin{equation}
\int d\alpha S_{\mathrm{toy,av}}(x)=
2\pi{\rm e}^{-\frac{r^2}{R^2}}I_0\left(\epsilon_2 \frac{r^2}{R^2}\right) I_0\left(\epsilon_3 \frac{r^2}{R^2}\right).
\end{equation}
However, if we had set $\epsilon_3=0$ and then integrated over $\alpha$, we would have obtained
\begin{equation}
\int d\alpha S_{\mathrm{toy}}(x;\epsilon_3=0)=
2\pi\exp\left[-\frac{r^2}{R^2}\right]I_0\left(\epsilon_2 \frac{r^2}{R^2}\right)\,  .
\end{equation}
The two results differ by a factor of $I_0\left(\epsilon_3 {r^2}/{R^2}\right)$, even in
this very simple case, and in the more complicated case of Fig.~\ref{fig:raver}.
It is also clear from this that averaging over random variations between
the difference of the third order and second order event planes, or assuming that there
are no third order variations in the density profile leads to very similar results for $\epsilon_3\ll1$,
with corrections of the order of $\epsilon_3^2$. It also turns out, that the same is true for third order
oscillations: event-plane averaged second order anisotropies have an effect of
the size $I_0\left(\epsilon_2 {r^2}/{R^2}\right)$.


\begin{figure}
\centering
\includegraphics[angle=-90,width=0.75\linewidth]{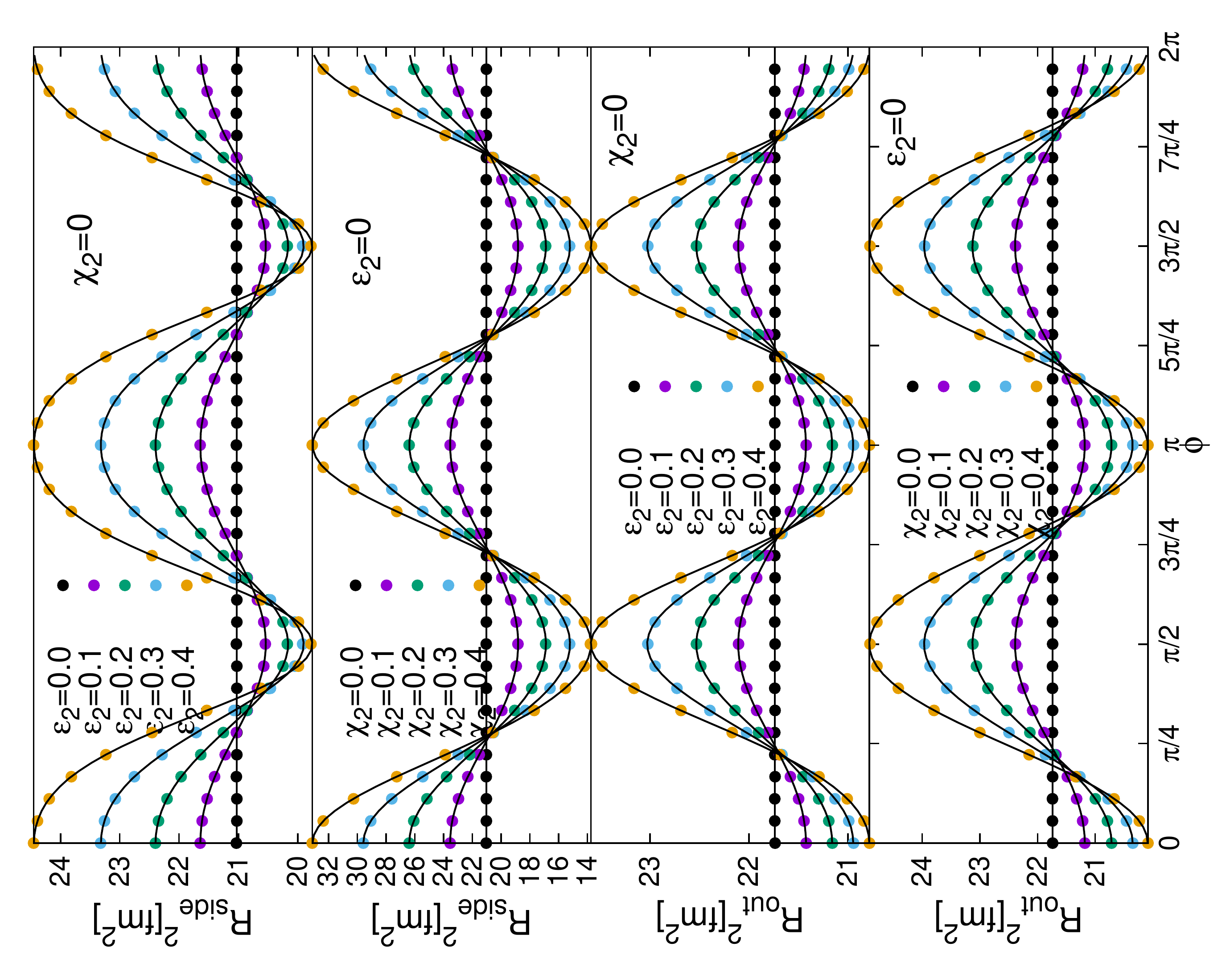}
\caption{\label{fig:oscR2} 
The $\phi$-dependence of $R_\out^2$ and $R_\side^2$ after setting $\psi_2=\alpha_2=0$ and
averaging over $\Delta\alpha_{23}$. Values of parameters are indicated in the panels. Points are 
results of calculations, lines are fits with Eq.~(\ref{eq:four2}).
}
\end{figure}

\begin{figure}
\centering
\includegraphics[angle=-90,width=0.75\linewidth]{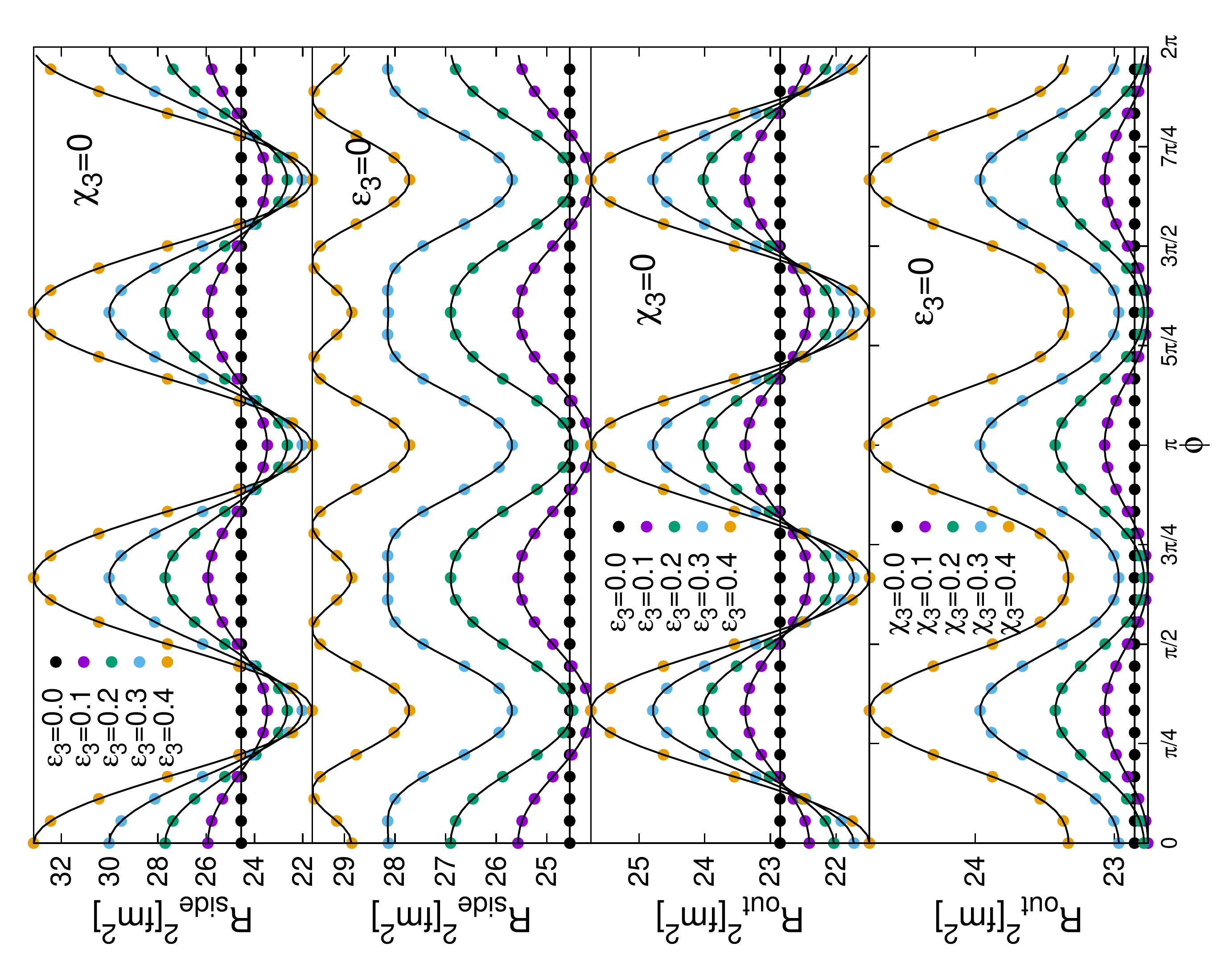}
\caption{\label{fig:oscR3} 
The $\phi$-dependence of $R_\out^2$ and $R_\side^2$ after setting $\psi_3=\alpha_3=0$ and
averaging over $\Delta\alpha_{23}$. Values of parameters are indicated in the panels. Points are 
results of calculations, lines are fits with Eq.~(\ref{eq:four3}).
}
\end{figure}

A numerical investigation of the effects of third order variations of the velocity profile
is indicated in Fig.~\ref{fig:raver}, that suggests that the relative error that comes from
the averaging over the random orientation of the third order event plane modifies the
amplitude of HBT oscillations slightly, and the modification increases with
increasing $\chi_3$, the coefficient of third order variations of the velocity profile.

In order to describe the azimuthal oscillations of Bose-Einstein or HBT radii, a blast-wave
model was also developed in ref.~\cite{Retiere:2003kf}. It was applied to study the second order
oscillations of pion and kaon HBT radii in ref.~\cite{Adare:2015bcj}. 
However, in these models, third order anisotropies as well as the
possible difference between the second order and third order event
planes have not been considered. Based on Fig.~\ref{fig:raver}, such an approximation may be
valid if $\xi_3$, the amplitude of third order oscillations 
in the local valocity distribution does not exceed the relative error of the experimental
determination of the HBT radii, corresponding to 5-10 \% in the case of a recently published
measurement of pion and kaon correlations~\cite{Adare:2015bcj}. 

This shows that if one needs to speed up the calculation then the easier way by setting 
some aniso\-tropies to 0 is viable. We have checked that this gives good results for the $\alpha_2=0$ 
leading Fourier order of the $\phi$-dependence and some deviations may appear (though not 
always) in the sub-leading terms. There are differences between the results of the two schemes for 
the case $\alpha_3=0$. The results presented here are obtained by conscientious averaging over 
$\Delta\alpha_{23}$.

In Fig.~\ref{fig:oscR2} we present the $\phi$-dependence of $R_\out^2$ and $R_\side^2$ with 
$\psi_2=\alpha_2=0$, for various values of $\epsilon_2$ and $\chi_2$ (while $\epsilon_3 = 0$, 
$\chi_3=0$). 
Similarly, in Fig.~\ref{fig:oscR3} the third-order oscillation is presented with $\psi_3=\alpha_3=0$ 
and second-order parameters set to $\epsilon_2=0$, and $\chi_2=0$. 
The common observation of these dependencies clearly shows that there are important 
next-to-leading order contributions in the Fourier expansions of $R_\out^2(\phi)$ and $R_\side^2(\phi)$.

\begin{figure}
\centering
\includegraphics[angle=-90,width=0.75\linewidth]{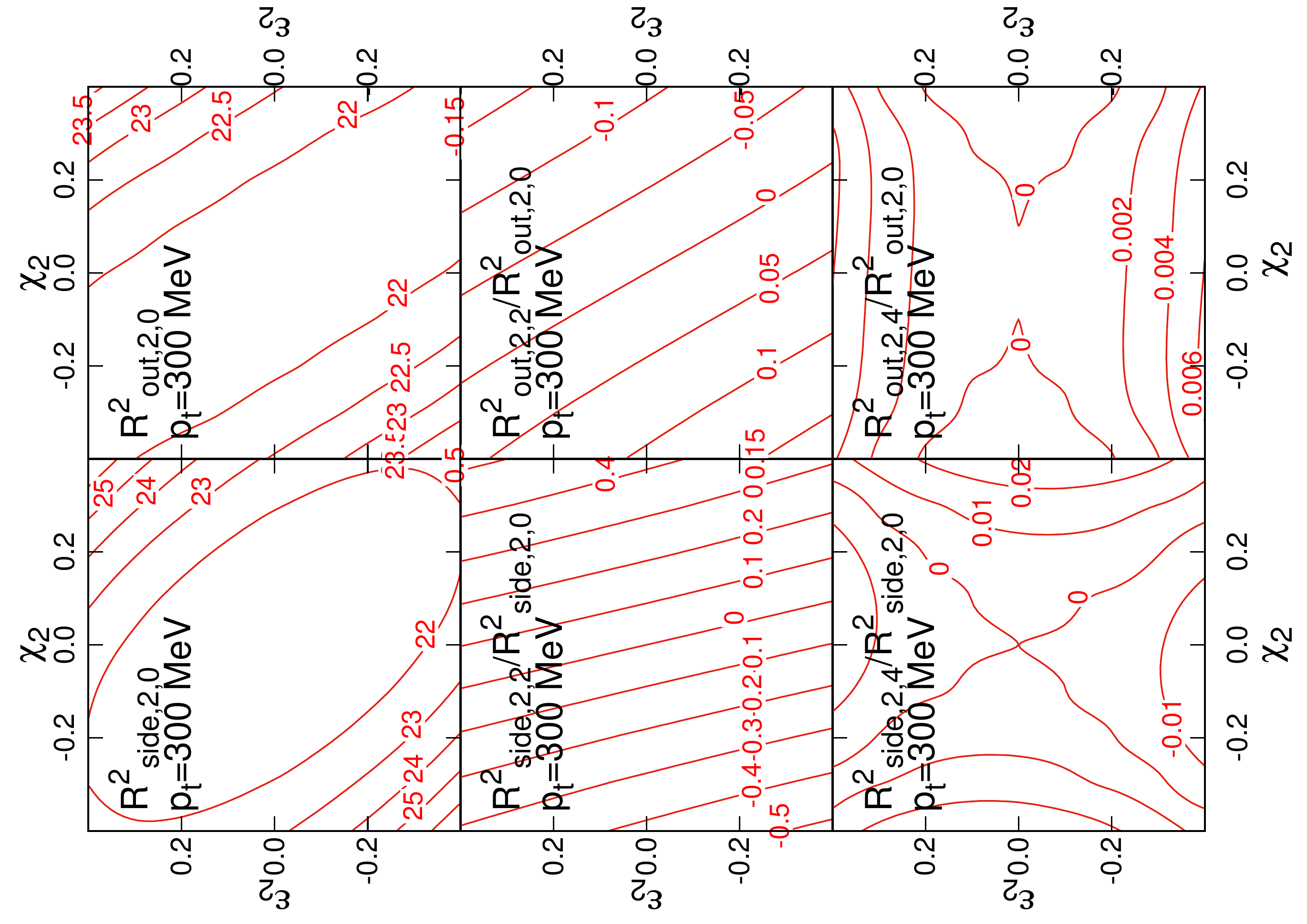}
\vspace{-5pt}
\caption{\label{fig:A2nd} 
Average radius and scaled amplitudes with $\psi_2=\alpha_2=0$ for $R_\side^2$ (left column) and $R_\out^2$ (right column)
average (top row), second (middle), and fourth (bottom) scaled amplitude, as functions of $\epsilon_2$ and $\chi_2$.
}
\end{figure}

\begin{figure}
\centering
\includegraphics[angle=-90,width=0.75\linewidth]{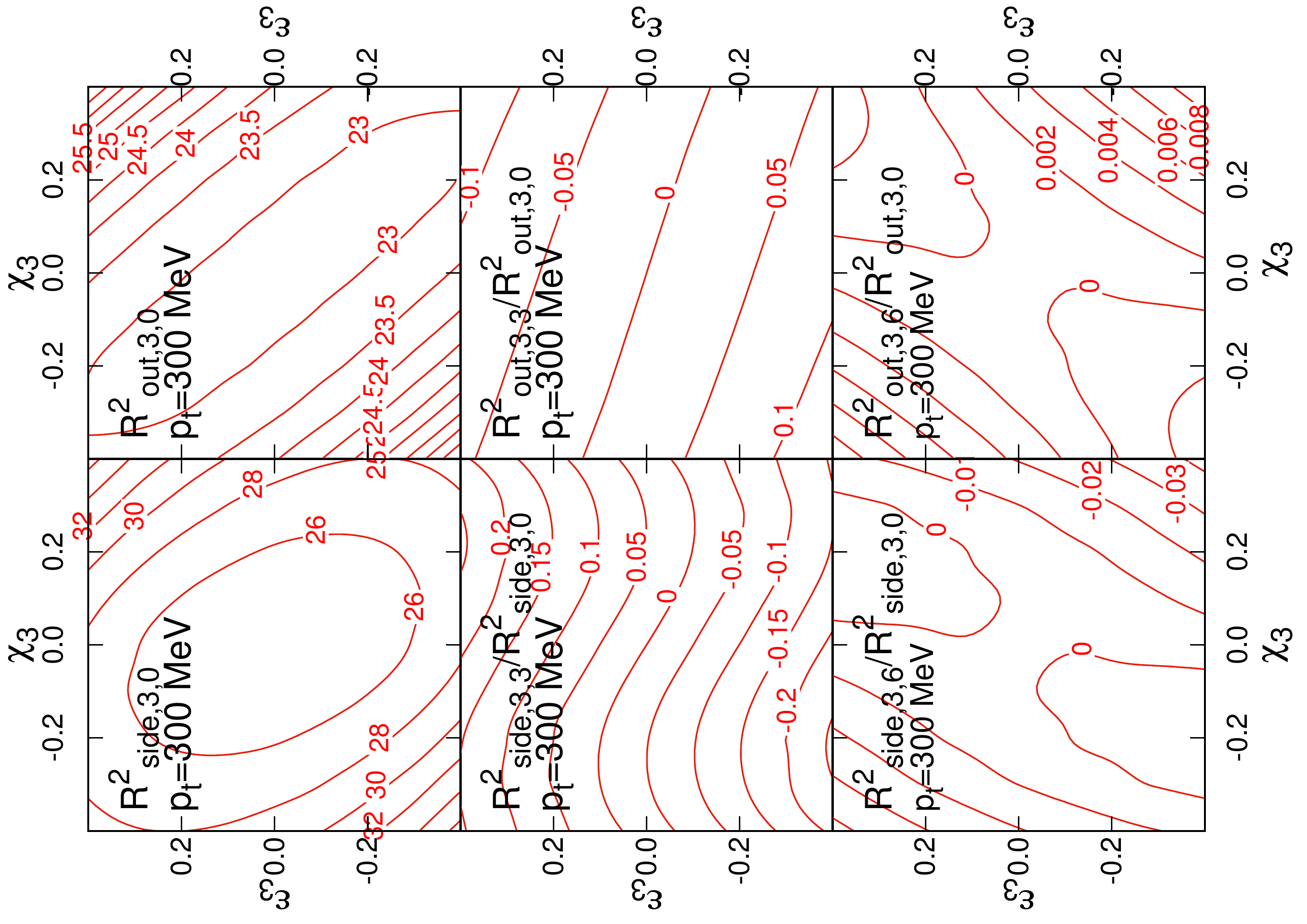}
\vspace{-5pt}
\caption{\label{fig:A3rd} 
Average radius and scaled amplitudes with $\psi_3=\alpha_3=0$ for $R_\side^2$ (left column) and $R_\out^2$ (right column)
average (top row), third (middle), and sixth (bottom) scaled amplitude, as functions of $\epsilon_2$ and $\chi_2$.
}
\end{figure}

To analyze this, the calculated values (data points) were fitted with Fourier series.
For $\psi_2=0$ this reads
\begin{align}
\label{eq:four2}
R_i^2(\phi) &= R_{i,2,0}^2 + R_{i,2,2}^2 \cos (2\phi) + R_{i,2,4}^2 \cos(4\phi) + R_{i,2,6}^2\cos(6\phi),
\end{align}
where $i$ stands for `out' or `side'. Higher-order terms have been neglected. For $\phi_3=0$ 
we have an expansion with terms of order 3 and its multiples
\begin{align}
\label{eq:four3}
R_i^2(\phi) &= R_{i,3,0}^2 + R_{i,3,3}^2 \cos (3\phi) + R_{i,3,6}^2 \cos(6\phi)  + R_{i,3,9}^2\cos(9\phi)\,  .
\end{align}
Note that in general $R_{i,3,6}^2 \neq R_{i,2,6}^2$. Therefore we have to introduce the cumbersome 
indexing of the Fourier terms which indicates the order of the event-plane set to 0 and the order of the term. 

The dependence of scaled amplitudes $R_{i,2,2}^2/R_{i,2,0}^2$ has been studied in 
Ref~\cite{Csanad:2008af}. Here we show it for completeness in Fig.~\ref{fig:A2nd}
together with the average radii and the scaled amplitudes for the fourth order, $R_{i,2,4}^2/R_{i,2,0}^2$.
Note the symmetry of the results with respect to the change $(\epsilon_2,\chi_2)\to(-\epsilon_2,-\chi_2)$. 
In fact, such a change is equivalent to a mere shift of the phase $\alpha_2$ by $\pi/2$. An interesting saddle-like
dependence is discovered for the fourth-order scaled amplitudes. While $R_{i,2,4}^2/R_{i,2,0}^2$
seems to vanish very roughly along the diagonals $\chi_2 = \pm \epsilon_2$, an important 
fourth order contribution shows up when one of the parameters is close to 0. 

For the $\alpha_3=\psi_3=0$ case we plot the average and the third and sixth-order scaled amplitudes as functions 
of $\chi_3$ and $\epsilon_3$ ($\epsilon_2=0$, $\chi_2=0$) in Fig.~\ref{fig:A3rd}.
Although we have kept the same $R$ in all calculations, the change in $R_{i,3,0}^2$ can be 
as large as 30\% for the investigated interval of $\epsilon_3$ and $\chi_3$. Smallest values 
are obtained roughly along $\epsilon_3 = -\chi_3$ which actually means that spatial and flow anisotropies 
have phases shifted by the maximum value of $\pi/6$. The largest radii are obtained for large values 
of $\epsilon_3=\chi_3$. 

The third-order scaled amplitude of $R_\side^2(\phi)$ shows an interesting dependence 
on $\chi_3$ and $\epsilon_3$. It seems to mainly depend on spatial anisotropy $\epsilon_3$,
so in first approximation $R_{\side,3,3}^2/R_{\side,3,0}^2$ could be used for the determination 
of $\epsilon_3$. However, we also observe a wavy structure if it is considered as a function of $\chi_3$,
as indicated by the second panel of Fig.~\ref{fig:oscR3}, so the 
statement holds only approximately. Nevertheless, we observe that the correlation  between 
$\chi_3$ and $\epsilon_3$ which leads to the same value of $R_{\out,3,3}^2/R_{\out,3,0}^2$
is almost perpendicular to that which yields the same $v_3$ (cf.~Fig.~\ref{fig:flows}). 
Thus from the combination of the two measurements one should be able to determine 
both third-order anisotropy parameters of this model. 

The contribution from the sixth-order Fourier coefficient has its gradient roughly along the 
line $\chi_3=-\epsilon_3$. This is the same direction as the one along which we observe the 
smallest average values of the correlation radii. 

\begin{figure}
\centering
\includegraphics[angle=-90,width=0.67\linewidth]{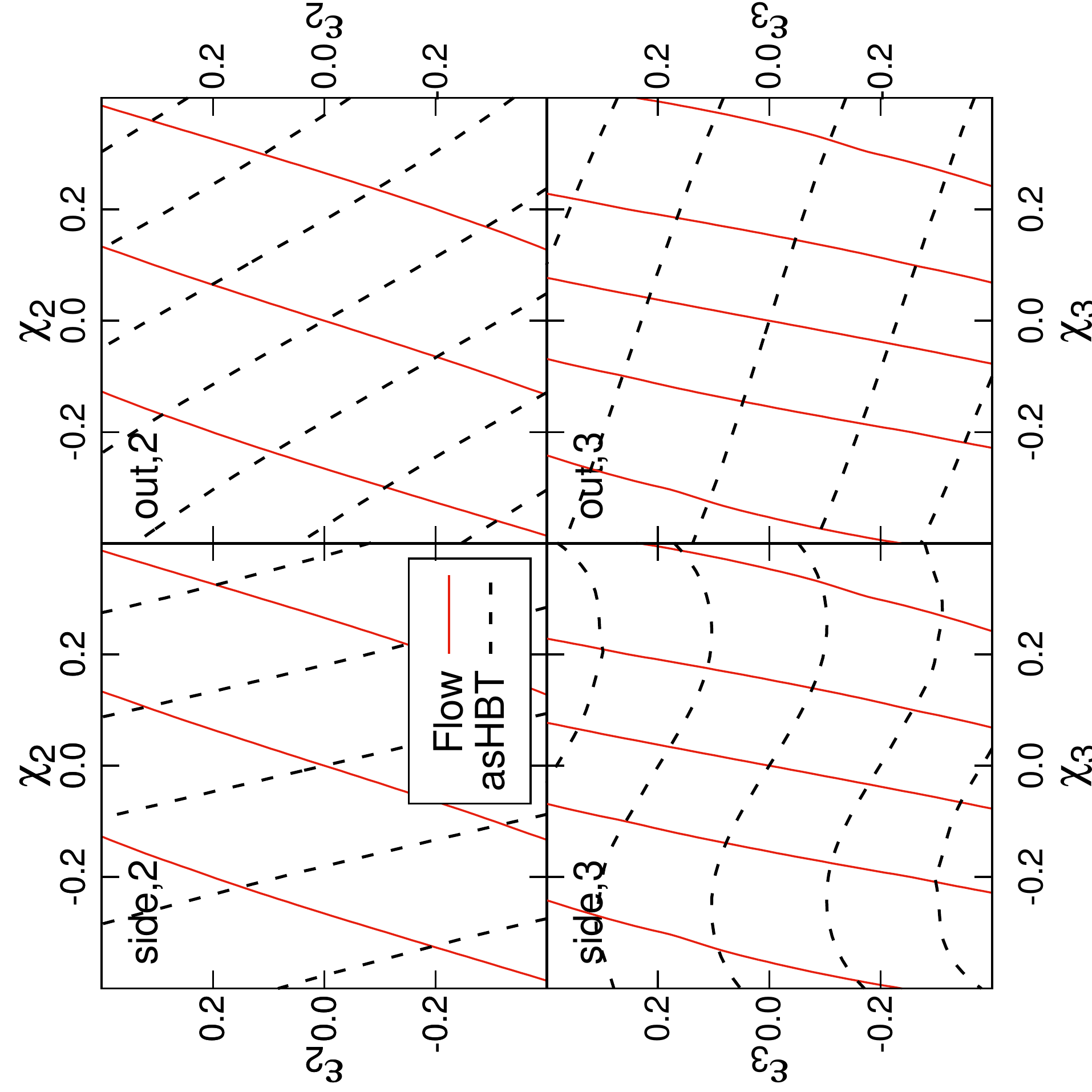}
\caption{\label{fig:disentangle} 
Flow coefficients and scaled amplitudes from Figs.~\ref{fig:flows},\ref{fig:A2nd}-\ref{fig:A3rd} superimposed,
as functions of $\epsilon_2$, $\chi_2$, $\epsilon_3$, and $\xi_3$.
The upper panels show the second order scaled HBT amplitudes as well as the elliptic flow, while the
lower panels show the third order scaled HBT amplitudes as well as the triangular flow. 
The lines that go through the point (0,0) in the $\epsilon,\chi$ plane correspond to the 0
value of the given observable. The values of the consecutive contours can be read out from
Figs.~\ref{fig:A2nd}-\ref{fig:A3rd}.
}
\end{figure}

The most important observation is however, that with the contours of Fig.~\ref{fig:flows} and
Figs.~\ref{fig:A2nd}-\ref{fig:A3rd}, indeed the spatial and flow-field anisotropies can be
disentangled. For illustration, in Fig.~\ref{fig:disentangle} we show how the particular values
of the oscillation of the correlation radii and of the flows let one determine the contribution
from $\epsilon_2$, $\chi_2$, $\epsilon_3$, and $\xi_3$.

\section{Conclusions}

We have extended the Buda-Lund hydro model with higher-order
anisotropies in transverse shape and expansion velocity profile. In a special case, this 
model can be identified with a solution of a certain hydrodynamic model. 

With the extended model we pushed further the study that has been started in~\cite{Csanad:2008af}. 
There, the influences of second-order anisotropies in space and expansion on the observable 
$v_2$ and the elliptic modulation of the correlation radii were investigated. It was deduced, how to disentangle 
them with the help of the following observations: to consider both the elliptic flow and the second order HBT oscillations.
In a similar manner, here we showed that the third-order anisotropies in space and expansion velocity 
can be disentangled if $v_3$ and $R^3_{i,3,3}/R^2_{i,3,0}$ 
are studied experimentally.

Within 
the Buda-Lund model we investigated how the mean value  of the correlation radii and the absolute 
normalization of single particle $p_t$ spectra increase if we average over the azimuthal angle. 

The conclusions drawn here were deduced from the results obtained with the extended 
Buda-Lund model. There are other analogical parameterizations of the freeze-out state 
of the fireball on the market, however. Examples are the Blast-wave model~\cite{Retiere:2003kf,Tomasik:2004bn} and/or
the Cracow single freeze-out model~\cite{Broniowski:2001we}. In the next future we 
therefore plan to implement higher-order anisotropies also in the Blast-wave model and perform a similar study. 
This will show which features of the results obtained here are robust and which are rather an artifact 
of the specific model. 

Future analytic investigations of solutions with small density and velocity perturbations
along the lines of ref.~\cite{Shi:2014kta} on the top of those exact hydrodynamical solutions
that form the basis of the Buda-Lund hydro model (for example
refs.~\cite{Csanad:2003qa,Csanad:2005gv})
are also among the future research directions that we consider important to pursue.

\vspace{-5pt}
\paragraph{Acknowledgments}
This research was partially supported by the Hungarian OTKA NK 101438 grant.
MCs is grateful for the support of the Hungarian American Enterprise Scholarship
Fund and by the J\'anos Bolyai Research Scholarship of the Hungarian Academy of Sciences.
LS thanks for the hospitality of BT in Bansk\'a Bystrica. BT acknowledges partial support
from VEGA 1/0469/15, APVV-0050-11 (Slovakia), and LG15001 (Czech Republic).


\vspace{-5pt}
\begin{thebibliography}{10}

\bibitem{Csorgo:1995bi}
T. Cs\"org\H{o} and B. L\"orstad, Phys. Rev. {\bf C54},  1390  (1996) [\href{http://www.arxiv.org/abs/hep-ph/9509213}{hep-ph/9509213}].

\bibitem{Csanad:2003qa}
M. Csan\'ad, T. Cs\"org\H{o}, and B. L\"orstad, Nucl. Phys. {\bf A742},  80  (2004) [\href{http://www.arxiv.org/abs/nucl-th/0310040}{nucl-th/0310040}].

\bibitem{Csanad:2008af}
M. Csan\'ad, B. Tom\'a\v{s}ik, and T. Cs\"org\H{o}, Eur. Phys. J. A {\bf 37},  111  (2008) [\href{http://www.arxiv.org/abs/0801.4434}{0801.4434}].

\bibitem{Ster:2010ia}
A. Ster, M. Csan\'ad, T. Cs\"org\H{o}, and B. L\"orstad, and B. Tom\'a\v{s}ik, Eur.Phys.J. {\bf A47},  58  (2011) [\href{http://www.arxiv.org/abs/1012.5084}{arXiv:1012.5084}].

\bibitem{Adare:2011tg}
A. Adare {\it et~al.}, Phys.Rev.Lett. {\bf 107},  252301  (2011) [\href{http://www.arxiv.org/abs/1105.3928}{arXiv:1105.3928}].

\bibitem{Aamodt:2011by}
K. Aamodt {\it et~al.}, Phys. Lett. {\bf B708},  249  (2012) [\href{http://www.arxiv.org/abs/1109.2501}{arXiv:1109.2501}].

\bibitem{Adamczyk:2013waa}
L. Adamczyk {\it et~al.}, Phys.Rev. {\bf C88},  014904  (2013) [\href{http://www.arxiv.org/abs/1301.2187}{arXiv:1301.2187}].

\bibitem{Adare:2014vax}
A. Adare {\it et~al.}, Phys.Rev.Lett. {\bf 112},  222301  (2014) [\href{http://www.arxiv.org/abs/1401.7680}{arXiv:1401.7680}].

\bibitem{Csorgo:1994in}
T. Cs\"org\H{o}, B. L\"orstad, and J. Zim\'anyi, Z. Phys. {\bf C71},  491  (1996) [\href{http://www.arxiv.org/abs/hep-ph/9411307}{hep-ph/9411307}].

\bibitem{Csorgo:1999sj}
T. Cs\"org\H{o}, Acta Phys. Hung. Ser. A: Heavy Ion Phys. {\bf 15},  1  (2002) [\href{http://www.arxiv.org/abs/hep-ph/0001233}{hep-ph/0001233}].

\bibitem{Csanad:2005gv}
M. Csan\'ad {\it et~al.}, Eur. Phys. J. {\bf A38},  363  (2008) [\href{http://www.arxiv.org/abs/nucl-th/0512078}{nucl-th/0512078}].

\bibitem{Cooper:1974mv}
F. Cooper and G. Frye, Phys. Rev. {\bf D10},  186  (1974).

\bibitem{Csorgo:2003ry}
T. Cs\"org\H{o}, L.~P. Csernai, Y. Hama, and T. Kodama, Acta Phys. Hung. Ser. A: Heavy Ion Phys. {\bf  A21},  73  (2004).

\bibitem{Csanad:2014dpa}
M. Csan\'ad and A. Szab\'o, Phys.Rev. {\bf C90},  054911  (2014) [\href{http://www.arxiv.org/abs/1405.3877}{arXiv:1405.3877}].

\bibitem{Csanad:2004mm}
M. Csan\'ad, T. Cs\"org\H{o}, B. L\"orstad, and A. Ster, J. Phys. {\bf G30},  S1079  (2004) [\href{http://www.arxiv.org/abs/nucl-th/0403074}{nucl-th/0403074}].

\bibitem{Makhlin:1987gm}
A.~N. Makhlin and Y.~M. Sinyukov, Z. Phys. {\bf C39},  69  (1988)	.

\bibitem{Kopecna:2015fwa}
R. Kope\v{c}n\'a and B. Tom\'a\v{s}ik, Eur. Phys. J. A {\bf 52}, 115 (2016) [\href{http://www.arxiv.org/abs/1506.06776}{arXiv:1506.06776}].

\bibitem{Retiere:2003kf}
F. Reti\`ere and M.~A. Lisa, Phys. Rev. {\bf C70},  044907  (2004) [\href{http://www.arxiv.org/abs/nucl-th/0312024}{nucl-th/0312024}].

\bibitem{Adare:2015bcj}
A. Adare {\it et~al.}, Phys. Rev. {\bf C92},  034914  (2015) [\href{http://www.arxiv.org/abs/1504.05168}{arXiv:1504.05168}].

\bibitem{Tomasik:2004bn}
B. Tom\'a\v{s}ik, Acta Phys. Polon. {\bf B36},  2087  (2005) [\href{http://www.arxiv.org/abs/nucl-th/0409074}{nucl-th/0409074}].

\bibitem{Broniowski:2001we}
W. Broniowski and W. Florkowski, Phys. Rev. Lett. {\bf 87},  272302  (2001) [\href{http://www.arxiv.org/abs/nucl-th/0106050}{nucl-th/0106050}].

\bibitem{Shi:2014kta}
S. Shi, J. Liao, and P. Zhuang, Phys. Rev. {\bf C90},  064912  (2014) [\href{http://www.arxiv.org/abs/1405.4546}{arXiv:1405.4546}].

\end{thebibliography}
\end{document}